\documentclass[aps, showpacs, showkeys,nofootinbib,floatfix]{revtex4}

\usepackage{amssymb}
\usepackage{amsmath}
\usepackage{graphicx}

\begin{document}

\title{Constraints on f(R) cosmologies from strong gravitational lensing systems}

\author{Kai Liao, Zong-Hong Zhu}
 \email{zhuzh@bnu.edu.cn}
\affiliation{Department of Astronomy, Beijing Normal University,
Beijing 100875, China}

\begin{abstract}
f(R) gravity is thought to be an alternative to dark energy which can explain the acceleration of
the universe. It has been tested by different observations including type Ia supernovae (SNIa), the
cosmic microwave background (CMB), the baryon acoustic oscillations (BAO) and so on. In this Letter,
we use the Hubble constant independent ratio between two angular diameter distances $D=D_{ls}/D_s$ to
constrain f(R) model in Palatini approach $f(R)=R-\alpha H^2_0(-\frac{R}{H^2_0})^\beta$. These data are from
various large systematic lensing surveys and lensing by galaxy clusters combined with X-ray observations.
We also combine the lensing data with CMB and BAO, which gives a stringent constraint.
The best-fit results are $(\alpha,\beta)=(-1.50,0.696)$ or $(\Omega_m,\beta)=(0.0734,0.696)$
using lensing data only. When combined with CMB and BAO,
the best-fit results are $(\alpha,\beta)=(-3.75,0.0651)$ or $(\Omega_m,\beta)=(0.286,0.0651)$.
If we further fix $\beta=0$ (corresponding to $\Lambda$CDM), the best-fit value for $\alpha$ is
$\alpha$=$-4.84_{-0.68}^{+0.91}(1\sigma)_{-0.98}^{+1.63}(2\sigma)$ for the lensing analysis and
$\alpha$=$-4.35_{-0.16}^{+0.18}(1\sigma)_{-0.25}^{+0.3}(2\sigma)$ for the combined data, respectively.
Our results show that $\Lambda$CDM model is within 1$\sigma$ range.

\end{abstract}


\maketitle

\section{$\text{Introduction}$}
One of the most striking things in modern cosmology is the universe undergoing an accelerated state  \cite{accelertion}.
In order to explain this phenomenon, people have introduced new component which is known as dark energy.
The simplest model is cosmological constant ($\Lambda$CDM). It is consist with all kinds of observations while it
indeed encounters the coincidence problem and the "fine-tuning" problem. Besides, there are many other dark energy
models including holographic dark energy \cite{holographic}, quintessence \cite{quintessence}, quintom \cite{quintom}, phantom \cite{phantom}, generalized Chaplygin
gas \cite{GCG} and
so on.
Besides dark energy, the acceleration can be explained in other ways. If the new component with negative pressure does not exist,
General Relativity (GR) should be modified. Until now, at least two effective theories have been proposed. One is considering
the extra dimensions which is related to the brane-world cosmology \cite{DGP}. The other is the so-called f(R) gravity \cite{f(R)}.
It changes the form of Einstein-Hilbert Lagrangian by f(R) expression. These theories can give an acceleration solution naturally
without introducing dark energy.
There are two kinds of forms about the f(R), the metric and the Palatini formalisms \cite{formalisms}. They give different dynamical equations. They can be unified only in the case of linear action (GR).
For the Palatini approach, the form $f(R)=R-\alpha H^2_0(-\frac{R}{H^2_0})^\beta$ is chosen so that it can result in the
radiation-dominated, matter-dominated and recent accelerating state. Furthermore, it can pass the solar system and has the correct
Newtonian limit \cite{Newtonian}.
In this Letter, we consider the Palatini formalisms. Under this assumption, the f(R) cosmology has two parameters. What we want
to emphasize is, among the parameters $(\alpha,\beta,\Omega_m)$, only two of them are independent. Therefore, we can exhibit the constraint
results on either $(\alpha,\beta)$ space or $(\Omega_m,\beta)$ space.
Various observations have already been used to constrain f(R) gravity including SNIa, CMB, BAO, Hubble parameter (H(z)) and so on.
Among these works, parameter $\beta$ has been constrained to very small value. In these papers \cite{constraint1},
they get $\beta\sim 10^{-1}$; in \cite{constraint2}, the matter power spectrum from the SDSS
gives $\beta\sim 10^{-5}$; in \cite{constraint3}, the $\beta$ was constrained to $\sim 10^{-6}$.
From these results, the f(R) gravity seems hard to be distinguished from the standard theory, where $\beta=0$. One effective way to solve this problem
in astronomy is combining different cosmological probes.
Strong lensing has been used to study both cosmology \cite{lensing1} and galaxies including
their structure, formation and evolution \cite{lensing2}.
The observations of the images combined with lens models can give us the information about the ratio between two angular diameter
distances, $D_{ls}$ and $D_s$. The former one is the distance between lens and source, the latter one
is the distance from observer to the source. Because the angular diameter distance depends
on cosmology, the $D_{ls}/D_s$ data can be used to constrain the parameters
in f(R) gravity. In this Letter, we select 63 strong lensing systems from SLACS
and LSD surveys assuming the singular isothermal sphere (SIS) model or the singular isothermal ellipsoid (SIE) model is right. Moreover, a sample of 10 giant
arcs is also contained. Using these 73 data, we try to give a new approach to constraining f(R) gravity.

This Letter is organized as follows. In Section 2, we briefly describe the basic theory about f(R) gravity and
the corresponding cosmology. In Section 3, we introduce the lensing data we use, the CMB data and the BAO data.
The constraint results are performed in Section 4. At last, we give a summary in Section 5. Throughout this work,
the unit with light velocity $c=1$ is used.

\section{$\text{The f(R) gravity and cosmology}$}
The basic theory of f(R) gravity has been discussed thoroughly in history. For details, see Ref. \cite{f(R)}.
In Palatini approach, the action is given by
\begin{equation}
S=-\frac{1}{2\kappa}\int{d^4x\sqrt{-g}f(R)}+S_m,
\end{equation}
where $\kappa=8\pi G$, $G$ is the gravitational constant and $S_m$ is the usual action for the matter.
The Ricci scalar depends on the metric and the affine connection:
\begin{equation}
R=g^{\mu\nu}\hat{R}_{\mu\nu},
\end{equation}
where the generalized Ricci tensor
\begin{equation}
\hat{R}_{\mu\nu}={\hat{\Gamma}^\alpha}_{\mu\nu,\alpha}-{\hat{\Gamma}^\alpha}_{
\mu\alpha,\nu}+{\hat{\Gamma}^\alpha}_{\alpha\lambda}{\hat{\Gamma}^\lambda}_{
\mu\nu}-{\hat{\Gamma}^\alpha}_{\mu\lambda}{\hat{\Gamma}^\lambda}_{\alpha\nu}.
\end{equation}
The hat represents the affine term which is different from the Levi-Civita connection. The Ricci
scalar is always negative. By varying the action with respect to the metric components, we can get
the generalized Einstein field equations:
\begin{equation}
f'(R)\hat{R}_{\mu\nu}-\frac{1}{2}g_{\mu\nu}f(R)=-\kappa
T_{\mu\nu},
\label{field Eq}
\end{equation}
where $f'(R)=df/dR$ and $T_{\mu\nu}$ is the matter energy-momentum tensor. For a perfect fluid,
$T_{\mu\nu} = (\rho_m + p_m)u_{\mu}u_{\nu} + p_m g_{\mu\nu}$,
where $\rho_m$ is the energy density, $p_m$ is the pressure and $u_{\mu}$ is the four-velocity.
Varying the action with respect to the connection gives the equation
\begin{equation}
\hat{\nabla}_\alpha[f'(R)\sqrt{-g}g^{\mu\nu}]=0.
\end{equation}
From this equation, we can obtain a conformal metric
$\gamma_{\mu\nu} = f'(R)g_{\mu\nu}$
which is corresponding to the affine connection.
The generalized Ricci tensor can be related to the Ricci tensor
\begin{equation}
\hat{R}_{\mu\nu}=R_{\mu\nu}-\frac{3}{2}\frac{\nabla_\mu f'\nabla_\nu
f'}{f'^2}+\frac{\nabla_\mu\nabla_\nu
f'}{f'}+\frac{1}{2}g_{\mu\nu}\frac{\nabla^\mu\nabla_\mu f'}{f'}.
\label{ricci tensor}
\end{equation}
In the next, we will introduce the dynamical equations of f(R) cosmology. Since all kinds of observations
support a flat universe, we assume a flat FRW cosmology. The FRW metric is
\begin{equation}
ds^2=-dt^2+a(t)^2\delta_{ij}dx^idx^j,
\end{equation}
where the scale factor $a=(1+z)^{-1}$, $z$ is the redshift. We choose $a_0=1$, the subscript "0" represents the
quantity today. From Eq.(\ref{ricci tensor}), we can obtain the generalized Friedmann equation
\begin{equation}
6(H+\frac{1}{2}\frac{\dot{f'}}{f'})^2=\frac{\kappa(\rho+3p)}{f'}-\frac{f}{f'},
\label{gfriedmann}
\end{equation}
where the overdot denotes a time derivative. The trace of Eq.(\ref{field Eq}) can gives
\begin{equation}
Rf'(R)-2f(R)=-\kappa T.
\label{trace}
\end{equation}
Considering the equation of state of matter is zero, Eq.(\ref{trace}) can give the relation between matter density and
redshift
\begin{equation}
(1+z)^{-1}=(\kappa\rho_{m0})^\frac{1}{3}(Rf'-2f)^{-\frac{1}{3}}.
\label{mz}
\end{equation}
Also, considering the energy conservation equation, Eq.(\ref{trace}) can give
\begin{equation}
\dot{R}=-\frac{3H\rho_M}{Rf''(R)-f'(R)}.
\label{dotr}
\end{equation}
According to Eq.(\ref{trace}), Eq.(\ref{dotr}) and Eq.(\ref{gfriedmann}), we can get the Hubble quantity in term of $R$
\begin{equation}
H^2(R)=\frac{1}{6f'}\frac{Rf'-3f}{(1-\frac{3}{2}\frac{f''(Rf'-2f)}{f'(Rf''-f')}
)^2}.
\label{hubble}
\end{equation}
This is the Friedmann equation in f(R) cosmology. For each $R$, we can get the redshift corresponding to that time.
The angular diameter distance between redshifts $z_1$ and $z_2$ is
\begin{eqnarray}
D^A(z_1,z_2)&=&\frac{1}{1+z_2}\int^{z_2}_{z_1}{\frac{dz}{H(z)}}
\\ \nonumber
&=&\frac{1}{3}(Rf'-2f)^{-\frac{1}{3}}\int^{R_{z_2}}_{R_{z_1}}{\frac{Rf''-f'}{
(Rf'-2f)^\frac{2}{3}}\frac{dR}{H(R)}}
\\ \nonumber
&=&D^A(R_1,R_2).
\end{eqnarray}
The $D_{ls}/D_s$ is given by
\begin{equation}
D_{ls}/D_s(z_1,z_2)=\frac{\int^{R_{z_2}}_{R_{z_1}}{\frac{Rf''-f'}{
(Rf'-2f)^\frac{2}{3}}\frac{dR}{H(R)}}}{\int^{R_{z_2}}_{R_{0}}{\frac{Rf''-f'}{
(Rf'-2f)^\frac{2}{3}}\frac{dR}{H(R)}}}.
\end{equation}

\section{$\text{Data and analysis methods}$}
In this section, we introduce the data we use, the lensing data, CMB and BAO. These data are independent of the Hubble constant.
\subsection{The $D_{ls}/D_s$ data}
Similar to Ref. \cite{cao}, our data set consists of two parts. Firstly, we choose 63 strong lensing systems from SLACS and LSD surveys \cite{lensingdata}. These systems have been measured the central dispersions with spectroscopic method. Though some of the lensing systems have 4 images, we assume the SIS or the SIE model is correct. The Einstein radius can be obtained under this assumption
\begin{equation}\label{ringeq}
\theta_E=4 \pi \frac{D_A(z,z_s)}{D_A(0,z_s)}
                 \frac{\sigma_{SIS}^2}{c^2}.
\end{equation}
It is related to the angular diameter distance ratio and stellar velocity dispersion $\sigma_{SIS}$  or the central velocity dispersion $\sigma_{0}$ which can be obtained from spectroscopy.
Secondly, the galaxy clusters can produce giant arcs, a sample of 10 galaxy clusters with redshift ranging from 0.1 to 0.6 is used under the $\beta$ model \cite{lensing4}.
Now, we have a sample of 73 strong lensing systems. There are listed in Table 2. We can fit the f(R) cosmology by minimizing the $\chi^{2}$ function
\begin{equation}\label{chi} \chi^2(\textbf{p})=
\sum_{i}\frac{(\mathcal{D}_i^{th}(\mathrm{\textbf{p}})-\mathcal{D}_{i}^{obs})^{2}}{\sigma
_{\mathcal{D},i}^{2}}.
\end{equation}

\subsection{Cosmic microwave background and baryon acoustic oscillation}
For CMB, the shift parameter $\cal R$ is an important quantity which depends on the cosmology \cite{cmb1}. In f(R) cosmology, it can be expressed as
\begin{eqnarray} \label{shift}
{\cal R} & = & \sqrt{\Omega_m H_0^2}\int_{0}^{z_{dec}}\frac{dz}{H(z)}\nonumber\\
& = & \sqrt{\Omega_m H_0^2}\int_{R_{dec}}^{R_0}\frac{a'(R)}{a(R)^2}
\frac{dR}{H(R)}\\
& = & \frac{1}{3^{4/3}}\left(\Omega_m H_0^2\right)^{1/6}\int_{R_0}^{R_{dec}}
\frac{Rf''-f'}{\left(Rf'-2f\right)^{2/3}}\frac{dR}{H(R)}\nonumber,
\end{eqnarray}
where $z_{dec}=1091.3$ is the redshift of the recombination epoch. The 7-year
WMAP gives the value ${\cal R}=1.725\pm 0.018$ \cite{cmb2}. The $\chi^2$ can be defined as
\begin{equation}
\chi^2_{CMB}=\frac{({\cal R}-1.725)^2}{0.018^2}.
\end{equation}

For BAO, we take the A parameter which is expressed as \cite{bao}
\begin{equation}
A= \sqrt{\Omega_m}E(z_{BAO})^{-1/3}\left[ \frac{1}{z_{BAO}}\int_{0}^{z_{BAO}}\frac{dz}{E(z)}\right]^{2/3},
\end{equation}
where $E(z)=H(z)/H_0$. The SDSS BAO measurement gives $A_{obs}=
0.469(n_s/0.98)^{-0.35} \pm 0.017$, where the scalar spectral index is
taken to be $n_s = 0.963$ as measured by WMAP7 \cite{cmb2}. The $\chi^2$ for BAO can be defined as
\begin{equation}
\label{chi2BAO} \chi_{BAO}^{2} = \frac {(A-A_{\rm
obs})^2}{\sigma^2_A}.
\end{equation}

\section{$\text{The constraint results}$}
In the Friedmann equation [12], we can find the Ricci scalar R is always divided by $H_0^2$, so we can choose units so that $H_0=1$. For given
$(\alpha,\beta)$, we can get the Ricci scalar today $R_0$ using the Friedmann equation. Then we can get $\Omega_m$ through Eq. [10]. Now, we can get
the relation between the Ricci scalar and the redshift through Eq. [10].
We use the 73 $D_{ls}/D_s$ data to constrain f(R) gravity in Palatini approach.
Fist, we show the $(\alpha,\beta)$ parameter space in Figure 1. We can see the
$D_{ls}/D_s$ data is compatible with the H(z) data \cite{hz}. The best-fit values are $(\alpha,\beta)=(-1.50,0.696)$.
 Using the $D_{ls}/D_s$ data
only cannot give a stringent constraint. After adding the CMB and the BAO data,
the parameters are tightly constrained. The best-fit values are $(\alpha,\beta)=(-3.75,0.0651)$. What we want to emphasis is the Hubble
parameter should always be positive, which restricts the parameters further.
We also exhibit the $(\Omega_m,\beta)$ parameter space in Figure 2. The best-fit values are $(\Omega_m,\beta)=(0.0734,0.696)$ for $D_{ls}/D_s$ data
and $(\Omega_m,\beta)=(0.286,0.0651)$ for combination with CMB and BAO.
Moreover, if we further fix $\beta=0$, the best-fit value for $\alpha$ is
$\alpha$=$-4.84_{-0.68}^{+0.91}(1\sigma)_{-0.98}^{+1.63}(2\sigma)$ for lensing data and
$\alpha$=$-4.35_{-0.16}^{+0.18}(1\sigma)_{-0.25}^{+0.3}(2\sigma)$ for combined data respectively.
From the results above, we can see the $\Lambda$CDM model which is corresponding to $(\alpha=-4.38,\beta=0)$ or $(\Omega_m=0.27,\beta=0)$ is within
$i\sigma$ range.
In order to compare the $D_{ls}/D_s$ data, we list some constraint results from other cosmological observations in Table 1.

\begin{figure}[t]
\centering
  \includegraphics[angle=0,width=150mm]{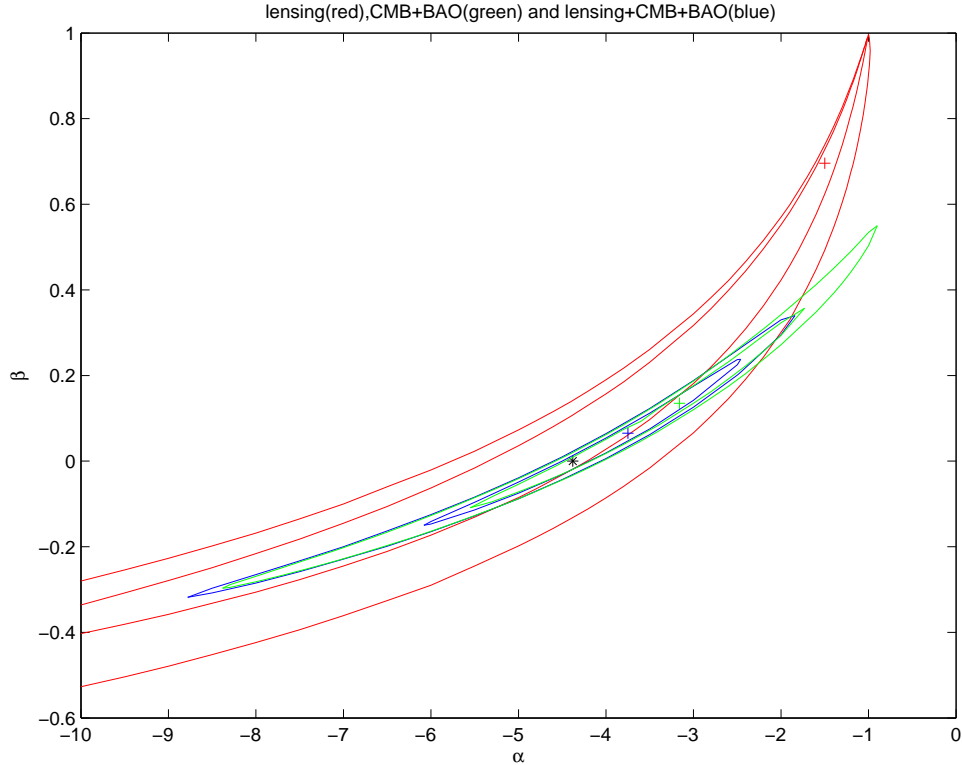}

\caption{
The $1\sigma$ and $2\sigma$ contours for $(\alpha,\beta)$ parameter space arising from the $D_{ls}/D_s$ data (red line), \textbf{CMB+BAO(green line)} and $D_{ls}/D_s$ data+CMB+BAO (blue line).
We have considered the parameter space that is not allowed. The black star represents the $\Lambda$CDM model $(\alpha=-4.38,\beta=0)$.
} \label{fig:phiCDM_Hz}
\end{figure}

\begin{figure}[t]
\centering
  \includegraphics[angle=0,width=150mm]{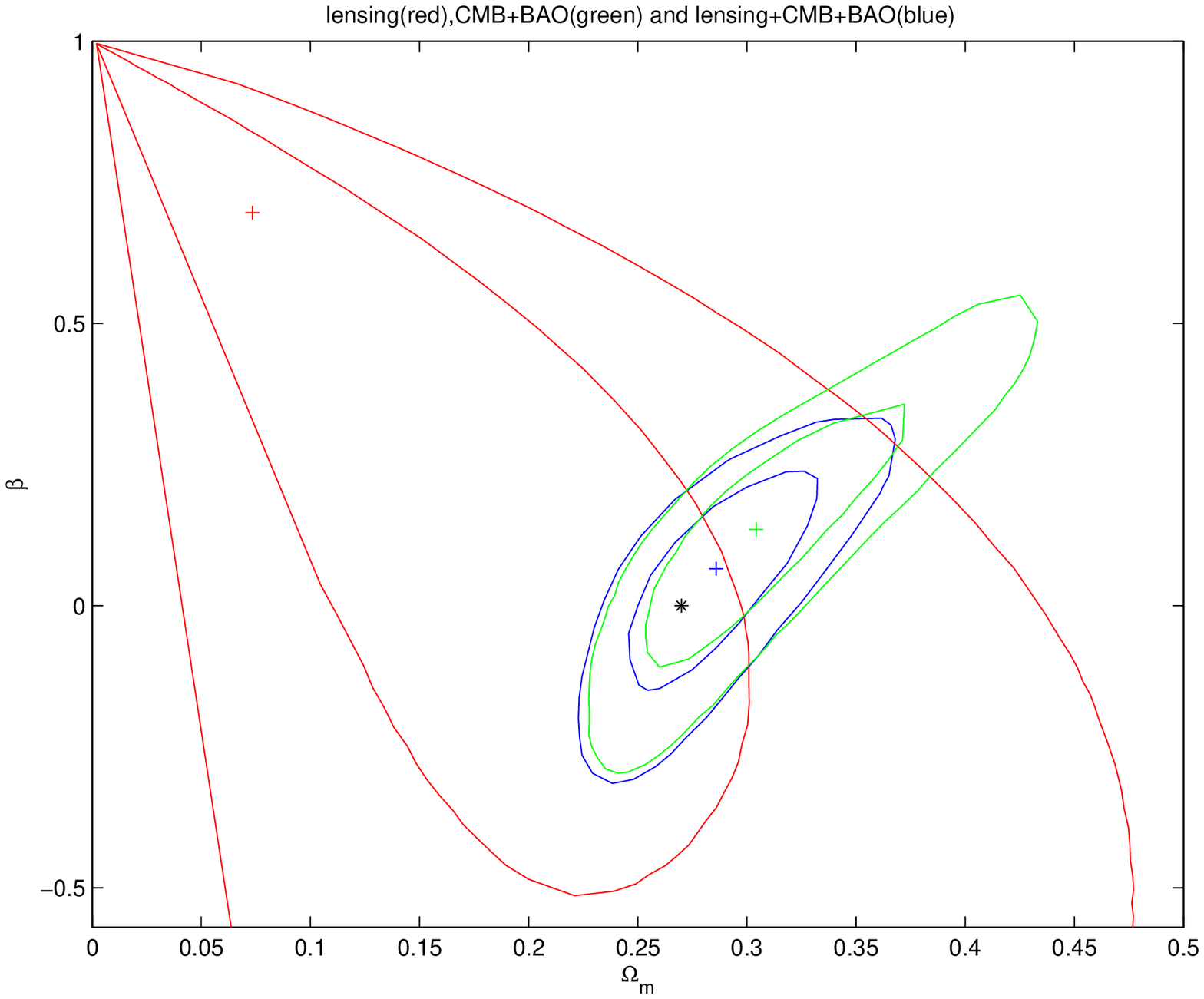}

\caption{
The $1\sigma$ and $2\sigma$ contours for $(\Omega_m,\beta)$ parameter space arising from the $D_{ls}/D_s$ data (red line), \textbf{CMB+BAO(green line)} and $D_{ls}/D_s$ data+CMB+BAO (blue line).
We have considered the parameter space that is not allowed. The black star represents the $\Lambda$CDM model $(\Omega_m=0.27,\beta=0)$.
} \label{fig:phiCDM_Hz}
\end{figure}

\begin{table*}[t]
\begin{center}
\begin{tabular}{lcrl}
\hline \hline \\
Test& Ref. & $\alpha$        &       $\beta$\\
\hline \hline \\
SNe Ia (SNLS) & \cite{constraintb} & -12.5 & -0.6\\
SNe Ia (SNLS) + BAO + CMB & \cite{constraintb} & -4.63 & -0.027\\
SNe Ia (Gold) & \cite{constrainta} & -10 & -0.51\\
SNe Ia (Gold) + BAO + CMB & \cite{constrainta} &  -3.6 &0.09\\
BAO & \cite{constrainta} & -1.1 & 0.57\\
CMB & \cite{constrainta} & -8.4 & -0.27\\
SNe Ia (Union) & \cite{constraintd} & - & -0.99\\
SNe Ia (Union) + BAO + CMB & \cite{constraintd} & -3.45 & 0.12\\
LSS & \cite{constraintc} &- & -2.6\\
H(z) & \cite{hz} &-1.11 & 0.9\\
H(z) + BAO + CMB& \cite{hz} & -4.7 & -0.03\\
Lensing($D_{ls}/D_s$)  & This Letter      & $-1.50_{-12.0}^{+0.52}$ & $0.696_{-1.21}^{+0.262}$\\
BAO+CMB  & This Letter  & $-3.16_{-2.39}^{+1.43}$  &  $0.135_{-0.244}^{+0.222}$\\
Lensing($D_{ls}/D_s$) + BAO + CMB & This Letter & $-3.75_{-2.33}^{+1.29}$ & $0.0651_{-0.2151}^{+0.1729}$\\

\hline \hline \\
\end{tabular}
\end{center}
\caption{ Best-fit values for $\alpha$ and $\beta$ (the $\Lambda$CDM model corresponds to $\alpha = -4.38$ and $\beta = 0$).}
\end{table*}

\section{$\text{Conclusion}$}
In this Letter, we use $D_{ls}/D_s$ data from lensing systems to constrain f(R) gravity in Palatini approach $f(R)=R-\alpha H^2_0(-\frac{R}{H^2_0})^\beta$.
Compared with references, we can see the constraint effects that $D_{ls}/D_s$ data give can be compatible with other data (SNe Ia, H(z), BAO, CMB and so on).
Moreover, we find although the best-fit values of the parameters are different from various observations, the directions of the contours in $(\alpha,\beta)$ space are very similar,
thus needing different observations to break the degeneracy. The $D_{ls}/D_s$ data propose a new way to probe the cosmology \cite{dlds}. As we expect,
the lensing data alone cannot give a stringent constraint. There are at least three aspects that contribute to the error. First, the assumption that
the lens galaxies satisfy SIS or SIE model may have some issues especially for four images. Second, the measurements of velocity dispersions have some
uncertainties. Finally, the error exists due to the influence of line of sight mass contamination \cite{sight}.
Combining with CMB and BAO, it gives $\beta\sim 10^{-1}$, which contains the
$\Lambda$CDM model. Until now, we cannot distinguish it from the standard cosmology, where $\beta=0$. For future lensing study, in order to improve the constraint,
we hope large survey projects can find more strong lensing systems. At the same time, a better understand about the lens model and more precise measurements can
give us more stringent results and more information about f(R) gravity.

\textbf{\ Acknowledgments } This work was supported by the National Natural Science Foundation
of China under the Distinguished Young Scholar Grant 10825313, the Ministry of Science and Technology national
basic science Program (Project 973) under Grant No.2012CB821804, the
Fundamental Research Funds for the Central Universities and
Scientific Research Foundation of Beijing Normal University.

\begin{table*}[ht]
\caption{We select 73 observations where the $D_{ls}/D_s<1$  from Ref. \cite{cao}  } \label{list}
\begin{center} {\tiny 
\begin{tabular}{lcccccccccc}

\hline
Cluster/galaxy & z$_s$ & z$_l$ & $\mathcal{D}^{obs}$ & $\sigma_\mathcal{D}$ \\
\hline
MS 0451.6-0305 & 2.91 & 0.550 & 0.785 & 0.087 \\
3C220.1 & 1.49 & 0.61 & 0.611 & 0.530 \\
CL0024.0 & 1.675 & 0.391 & 0.919 & 0.430  \\
Abell 2390 & 4.05 & 0.228 & 0.737 & 0.053  \\
Abell 2667 & 1.034 & 0.226 &  0.837 & 0.124 \\
Abell 68 & 1.6 & 0.255 & 0.982 & 0.225 \\
MS 1512.4 & 2.72 & 0.372 & 0.734 & 0.330 \\
MS 2137.3-2353 & 1.501 & 0.313 & 0.778 & 0.105 \\
MS 2053.7 & 3.146 & 0.583 & 0.968 & 0.209 \\
PKS 0745-191 & 0.433 & 0.103 & 0.818 & 0.065 \\
SDSS J0037-0942 & 0.6322 & 0.1955 & 0.6418 & 0.0501 \\
SDSS J0216-0813 & 0.5235 & 0.3317 & 0.3278  & 0.0451 \\
SDSS J0737+3216 & 0.5812 & 0.3223 & 0.3365  & 0.033 \\
SDSS J0912+0029 & 0.324  & 0.1642 & 0.5293 & 0.0391 \\
SDSS J0956+5100 & 0.47  & 0.2405 & 0.4532 & 0.0485 \\
SDSS J0959+0410 & 0.5349  & 0.126 & 0.6621  & 0.0752 \\
SDSS J1250+0523 & 0.795 & 0.2318 & 0.5319 & 0.0582 \\
SDSS J1330-0148 & 0.7115 & 0.0808 & 0.7762 & 0.0796 \\
SDSS J1402+6321 & 0.4814 & 0.2046 & 0.5739 & 0.0633 \\
SDSS J1420+6019 & 0.5352 & 0.0629 & 0.851 & 0.0413 \\
SDSS J1627-0053 & 0.5241 & 0.2076 & 0.4828 & 0.0426 \\
SDSS J1630+4520 & 0.7933 & 0.2479 & 0.8074 & 0.0984 \\
SDSS J2300+0022 & 0.4635 & 0.2285 & 0.4666 & 0.0581 \\
SDSS J2303+1422 & 0.517 & 0.1553 & 0.7754 & 0.0916 \\
SDSS J2321-0939 & 0.5324 & 0.0819 & 0.9082 & 0.0519 \\
Q0047-2808 & 3.595 & 0.485 & 0.8872 & 0.1162 \\
CFRS03-1077 & 2.941 & 0.938 & 0.6834 & 0.1035 \\
HST 14176 & 3.399 & 0.81 & 0.9757  & 0.1307 \\
HST 15433 & 2.092 & 0.497 & 0.929 & 0.1602 \\
MG 2016 & 3.263 & 1.004 & 0.5035 & 0.0982 \\
SDSS J0029-0055  &   0.9313  &   0.227   &   0.6356  &   0.0999  \\
SDSS J0044+0113  &   0.1965  &   0.1196  &   0.3877  &   0.0379  \\
SDSS J0109+1500  &   0.5248  &   0.2939  &   0.3803  &   0.0576  \\
SDSS J0330-0020  &   1.0709  &   0.3507  &   0.8498  &   0.1684  \\
SDSS J0728+3835  &   0.6877  &   0.2058  &   0.9477  &   0.0974  \\
SDSS J0822+2652  &   0.5941  &   0.2414  &   0.6056  &   0.0701  \\
SDSS J0841+3824  &   0.6567  &   0.1159  &   0.9671  &   0.0946  \\
SDSS J0935-0003  &   0.467   &   0.3475  &   0.1926  &   0.0341  \\
SDSS J0936+0913  &   0.588   &   0.1897  &   0.6409  &   0.0633  \\
SDSS J0946+1006  &   0.6085  &   0.2219  &   0.6927  &   0.1106  \\
SDSS J0955+0101  &   0.3159  &   0.1109  &   0.8571  &   0.1161  \\
SDSS J0959+4416  &   0.5315  &   0.2369  &   0.5599  &   0.0872  \\
SDSS J1016+3859  &   0.4394  &   0.1679  &   0.6204  &   0.0653  \\
SDSS J1020+1122  &   0.553   &   0.2822  &   0.524   &   0.0669  \\
SDSS J1023+4230  &   0.696   &   0.1912  &   0.836   &   0.1036  \\
SDSS J1029+0420  &   0.6154  &   0.1045  &   0.7952  &   0.0833  \\
SDSS J1032+5322  &   0.329   &   0.1334  &   0.4082  &   0.0414  \\
SDSS J1103+5322  &   0.7353  &   0.1582  &   0.9219  &   0.1129  \\
SDSS J1106+5228  &   0.4069  &   0.0955  &   0.6222  &   0.0617  \\
SDSS J1112+0826  &   0.6295  &   0.273   &   0.5052  &   0.0632  \\
SDSS J1134+6027  &   0.4742  &   0.1528  &   0.6687  &   0.0671  \\
SDSS J1142+1001  &   0.5039  &   0.2218  &   0.6967  &   0.1387  \\
SDSS J1143-0144  &   0.4019  &   0.106   &   0.8061  &   0.0779  \\
SDSS J1153+4612  &   0.8751  &   0.1797  &   0.7138  &   0.0948  \\
SDSS J1204+0358  &   0.6307  &   0.1644  &   0.6381  &   0.0813  \\
SDSS J1205+4910  &   0.4808  &   0.215   &   0.5365  &   0.0535  \\
SDSS J1213+6708  &   0.6402  &   0.1229  &   0.5783  &   0.0594  \\
SDSS J1403+0006  &   0.473   &   0.1888  &   0.6352  &   0.1014  \\
SDSS J1416+5136  &   0.8111  &   0.2987  &   0.8259  &   0.1721  \\
SDSS J1430+4105  &   0.5753  &   0.285   &   0.509   &   0.1012  \\
SDSS J1436-0000  &   0.8049  &   0.2852  &   0.775   &   0.1176  \\
SDSS J1443+0304  &   0.4187  &   0.1338  &   0.6439  &   0.0678  \\
SDSS J1451-0239  &   0.5203  &   0.1254  &   0.7262  &   0.0912  \\
SDSS J1525+3327  &  0.7173  &   0.3583  &   0.6526  &   0.1285  \\
SDSS J1531-0105  &   0.7439  &   0.1596  &   0.7628  &   0.0766  \\
SDSS J1538+5817  &   0.5312  &   0.1428  &   0.972   &   0.1234  \\
SDSS J1621+3931  &   0.6021  &   0.2449  &   0.8042  &   0.1363  \\
SDSS J1636+4707  &   0.6745  &   0.2282  &   0.7093  &   0.0921  \\
PG1115+080  &   1.72    &   0.31    &   0.7036  &   0.1252  \\
MG1549+3047 &   1.17    &   0.11    &   0.5728  &   0.0908  \\
Q2237+030   &   1.169   &   0.04    &   0.6685  &   0.1866  \\
CY2201-3201 &   3.9 &   0.32    &   0.8526  &   0.2624  \\
B1608+656   &   1.39    &   0.63    &   0.646   &   0.1831  \\
\hline

\end{tabular}}\\

\end{center}
\end{table*}
\end{document}